\documentclass[12pt]{article}
\usepackage[dvips]{color}
\usepackage{epsfig}
\usepackage{amsmath}
\usepackage{graphicx}

\begin{document}
\title{\bf Formulation of an Electrostatic Field with a Charge Density in the Presence of a Minimal Length Based on the Kempf Algebra}

\author{S. K. Moayedi $^{a}$\thanks{E-mail: s-moayedi@araku.ac.ir}\hspace{1mm}
, M. R. Setare $^{b}$ \thanks{E-mail:
rezakord@ipm.ir}\hspace{1mm},
 H. Moayeri $^{a}$\thanks{E-mail: h-moayeri@phd.araku.ac.ir}\hspace{1.5mm}  \\
$^a$ {\small {\em  Department of Physics, Faculty of Sciences,
Arak University, Arak 38156-8-8349, Iran}}\\
$^{b}${\small {\em Department of Science, Campus of Bijar,
University of Kurdistan Bijar, Iran
}}\\
}
\date{\small{}}
\maketitle
\begin{abstract}
\noindent

In a series of papers, Kempf and co-workers (J. Phys. A: Math.
Gen. {\bf 30}, 2093, (1997); Phys. Rev. D {\bf52}, 1108, (1995);
Phys. Rev. D {\bf55}, 7909, (1997)) introduced a D-dimensional
$(\beta,\beta')$-two-parameter deformed Heisenberg algebra which
leads to a nonzero minimal observable length. In this work, the
Lagrangian formulation of an electrostatic field in three spatial
dimensions described by Kempf algebra is studied in the case
where $\beta'=2\beta$ up to first order over deformation
parameter $\beta$. It is shown that there is a similarity between
electrostatics in the presence of a minimal length (modified
electrostatics) and higher derivative Podolsky's electrostatics.
The important property of this modified electrostatics is that
the classical self-energy of a point charge becomes a finite
value. Two different upper bounds on the isotropic minimal length
of this modified electrostatics are estimated. The first upper
bound will be found by treating the modified electrostatics as a
classical electromagnetic system, while the second one will be
estimated by considering the modified electrostatics as a quantum
field theoretic model. It should be noted that the quantum upper
bound on the isotropic minimal length in this paper is near to
the electroweak length scale $(\ell_{electroweak}\sim 10^{-18}\,
m)$.
\\

\noindent
\hspace{0.35cm}



\end{abstract}
\section{Introduction}
One of the fundamental problems in theoretical physics is the
unification between Einstein's general relativity and quantum
mechanics. The most interesting consequence of this unification is
the appearance of a minimal observable length on the order of the
Planck length. Today's theoretical physicists know that the
existence of a minimal observable length leads to a generalized
Heisenberg uncertainty principle. This generalized or
gravitational uncertainty principle (GUP) can be written as
\begin{equation}
\triangle X \triangle
P\geq\frac{\hbar}{2}\left[1+\alpha^{2}\ell_{P}^{2}\frac{(\triangle
P)^{2}}{\hbar^{2}}\right],
\end{equation}
where $\ell_{P}$ is the Planck length and $\alpha$ is a positive
numerical constant [1]. At low energies, the second term on the
right-hand side of Eq. (1) may be neglected
$(\alpha^{2}\ell_{P}^{2}\frac{(\triangle P)^{2}}{\hbar^{2}}\ll 1)$,
and we have $\triangle X \triangle P\geq\frac{\hbar}{2}$. The high
energy limit is obtained when
$\alpha^{2}\ell_{P}^{2}\frac{(\triangle P)^{2}}{\hbar^{2}}\sim 1$.
In this limit Eq. (1) leads to a minimal observable length
$(\triangle X)_{min}=\alpha\ell_{P}$. Many physicists believe that
introducing a minimal observable length into a quantum field theory
can eliminate the divergences [2-5]. In the past few years,
formulation of quantum field theory and gravity in the presence of a
minimal observable length have been studied extensively [6-19]. \\
This paper is organized as follows. In Sec. 2, the D-dimensional
$(\beta , \beta')$-two-parameter deformed Heisenberg algebra
introduced by Kempf is reviewed and it is shown that the above
algebra leads to a minimal observable distance [6-8]. In Sec. 3,
the Lagrangian formulation of an electrostatic field in three
spatial dimensions described by Kempf algebra is presented in the
case where $\beta' = 2\beta$ up to first order over deformation
parameter $\beta$. We show that there is a similarity between
electrostatics in the presence of a minimal length and higher
derivative Podolsky's electrostatics in three spatial dimensions.
The important property of electrostatics in the presence of a
minimal length is that the self-energy of a point charge becomes
a finite value. In Sec. 4, two upper bounds on the minimal length
of this modified electrostatics are obtained. The first upper
bound will be found by treating the modified electrostatics as a
classical electromagnetic system, while the second one will be
obtained by considering the modified electrostatics as a quantum
field theoretic model. Finally, in conclusions the numerical
values for the classical and quantum upper bounds on the
isotropic minimal length in this work are compared with the
results in previous investigations. SI units are used throughout
this paper.

\section{A Brief Review of the Kempf Algebra}
Let us begin with a brief review of the Kempf algebra, which is a
modified Heisenberg algebra that describes a D-dimensional quantized
space [6-8]. The D-dimensional Kempf algebra is specified by the
following modified commutation relations:
\begin{eqnarray}
\left[X_{i},P_{j}\right] &=& i\hbar \left[
\delta_{ij}(1+\beta\textbf{P}^{2}) +
\beta' P_{i}P_{j}\right], \\
\left[X_{i},X_{j}\right] &=& i\hbar
\frac{(2\beta-\beta')+\beta(2\beta+\beta')\textbf{P}^{2}}{1+\beta
\textbf{P}^{2}}(P_{i}X_{j}-P_{j}X_{i}), \\
\left[P_{i},P_{j}\right] &=& 0,
\end{eqnarray}

where $i,j=1,2,...,D$ and $\beta ,\beta'$ are two deformation
parameters which are nonnegative $(\beta,\beta' \geq 0)$ and have
the same dimensions, i.e., $[\beta]=[\beta']=(momentum)^{-2}$.
Also, $X_{i}$ and $P_{i}$ are position and momentum operators in
the deformed space. Using (2) and the Schwarz inequality for a
quantum state, the uncertainty relation for position and momentum
by assuming that $\triangle P_{i}$ is isotropic $(\triangle
P_{1}=\triangle P_{2}=...=\triangle P_{D})$ becomes [12]
\begin{equation}
(\triangle X_{i})(\triangle
P_{i})\geq\frac{\hbar}{2}\left[1+(D\beta+\beta')(\triangle
P_{i})^{2}+\gamma\right],
\end{equation}
where
$$ \gamma=\beta\sum_{k=1}^{D}<P_{k}>^{2}+\beta'<P_{i}>^{2}. $$
The relation (5) leads to an isotropic minimal length which is
given by
\begin{equation}
(\triangle X_{i})_{min}=\hbar\sqrt{D\beta+\beta'}\quad ,
\quad\forall i\in \{1,2, \cdots ,D\}.
\end{equation}
In Ref. 17, Stetsko and Tkachuk introduced a representation which
satisfies the modified commutation relations (2)-(4) up to first
order in deformation parameters $\beta , \beta'$. The
Stetsko-Tkachuk representation is given by
\begin{eqnarray}
X_{i} &=& x_{i}+
\frac{2\beta-\beta'}{4}(\textbf{p}^{2}x_{i}+x_{i}\textbf{p}^{2}), \\
P_{i} &=& p_{i}(1+\frac{\beta'}{2}\textbf{p}^{2}),
\end{eqnarray}
where $x_{i}$ , $p_{i}=\frac{\hbar}{i}\frac{\partial}{\partial
x_{i}}=\frac{\hbar}{i}\partial_{i}$ are position and momentum
operators in usual quantum mechanics, and
$\textbf{p}^{2}=\sum_{i=1}^{D}p_{i}^{2}$.\\
In this work, we only consider the special case $\beta'=2\beta$,
wherein the position operators commute to first order in $\beta$,
i.e., $[X_{i},X_{j}]=0$.\\
In such a linear approximation, the Kempf algebra reads
\begin{eqnarray}
\left[X_{i},P_{j}\right] &=&
i\hbar\left[\delta_{ij}(1+\beta\textbf{P}^{2})+2\beta
P_{i}P_{j}\right],\\
\left[X_{i},X_{j}\right] &=& 0,\\
\left[P_{i},P_{j}\right] &=& 0.
\end{eqnarray}
In Ref. 15, Brau showed that the following representations satisfy
(9)-(11), at the first order in $\beta$,
\begin{eqnarray}
X_{i} &=& x_{i},\\
P_{i} &=& p_{i}(1+\beta\textbf{p}^{2}).
\end{eqnarray}
Note that the representations (7) , (8) and (12) , (13) coincide
when $\beta' =2 \beta$.

\section{Lagrangian Formulation of an Electrostatic Field with a Charge Density Based on the Kempf Algebra}
The Lagrangian density for an electrostatic field with a charge
density $\rho (\textbf{x})$ in three spatial dimensions $(D=3)$ is
[20]
\begin{equation}
{\cal
L}(\phi,\partial_{i}\phi)=\frac{1}{2}\varepsilon_{0}(\partial_{i}\phi)(\partial_{i}\phi)-\rho\phi,
\end{equation}
where $\phi(\textbf{x})$ is the electrostatic potential. The
Euler-Lagrange equation for the electrostatic potential is
\begin{equation}
\frac{\partial{\cal
L}}{\partial\phi}-\partial_{i}\left(\frac{\partial{\cal
L}}{\partial(\partial_{i}\phi)}\right)=0.
\end{equation}
If we substitute the Lagrangian density (14) in the Euler-Lagrange
equation (15), we will obtain the Poisson equation as follows
\begin{equation}
\nabla^{2}\phi(\textbf{x})=-\frac{\rho(\textbf{x})}{\varepsilon_{0}},
\end{equation}
where $\nabla^{2}:=\partial_{i}\partial_{i}$ is the Laplace
operator. The Poisson equation (16) is equivalent to the following
two equations
\begin{eqnarray}
\nabla\cdot\textbf{E}(\textbf{x}) &=& \frac{\rho(\textbf{x})}{\varepsilon_{0}},\\
\nabla\times\textbf{E}(\textbf{x}) &=& 0,
\end{eqnarray}
where $\textbf{E}(\textbf{x})=-\nabla \phi(\textbf{x})$ is the
electrostatic field.\\
Now, let us obtain the Lagrangian density for an electrostatic field
in the presence of a minimal length based on the Kempf algebra. For
such a purpose, we must replace the usual position and derivative
operators $x_{i}$ and $\partial_{i}$ with the modified position and
derivative operators $X_{i}=x_{i}$ and $\nabla_{i}:=(1-\beta
\hbar^{2}\nabla^{2})\partial_{i}$ according to (12) and (13) in the
Lagrangian density (14), i.e.,
\begin{eqnarray}
{\cal L} &=&
\frac{1}{2}\varepsilon_{0}(\nabla_{i}\phi)(\nabla_{i}\phi)-\rho\phi\
\nonumber \\ &=& \frac{1}{2}\varepsilon_{0}[(1-\beta
\hbar^{2}\nabla^{2})\partial_{i}\phi][(1-\beta\hbar^{2}\nabla^{2})\partial_{i}\phi]-\rho\phi
\nonumber \\
 &=&
\frac{1}{2}\varepsilon_{0}(\partial_{i}\phi)(\partial_{i}\phi)+\varepsilon_{0}(\hbar\sqrt{\beta})^{2}(\partial_{i}\partial_{j}\phi)(\partial_{i}\partial_{j}\phi)
-\varepsilon_{0}(\hbar\sqrt{\beta})^{2}\partial_{j}[(\partial_{i}\phi)(\partial_{i}\partial_{j}\phi)]
\nonumber \\
 & & -\rho\phi+{\cal O}\left((\hbar\sqrt{\beta})^{4}\right).
\end{eqnarray}
After neglecting terms of order $(\hbar\sqrt{\beta})^{4}$ and
dropping out the total derivative term \\
$-
\varepsilon_{0}(\hbar\sqrt{\beta})^{2}\partial_{j}[(\partial_{i}\phi)(\partial_{i}\partial_{j}\phi)],$
the Lagrangian density (19) will be equivalent to the following
Lagrangian density
\begin{equation}
{\cal
L}(\phi,\partial_{i}\phi,\partial_{i}\partial_{j}\phi)=\frac{1}{2}\varepsilon_{0}(\partial_{i}\phi)(\partial_{i}\phi)
+\varepsilon_{0}(\hbar\sqrt{\beta})^{2}(\partial_{i}\partial_{j}\phi)(\partial_{i}\partial_{j}\phi)-\rho\phi.
\end{equation}
The term
$\varepsilon_{0}(\hbar\sqrt{\beta})^{2}(\partial_{i}\partial_{j}\phi)(\partial_{i}\partial_{j}\phi)$
in (20) can be considered as a minimal length effect. The
Euler-Lagrange equation for the generalized Lagrangian density (20)
is [21,22]
\begin{equation}
\frac{\partial{\cal
L}}{\partial\phi}-\partial_{i}\left(\frac{\partial{\cal
L}}{\partial(\partial_{i}\phi)}\right)+\partial_{i}\partial_{j}\left(\frac{\partial{\cal
L}}{\partial(\partial_{i}\partial_{j}\phi)}\right)=0.
\end{equation}
If we substitute (20) into (21), we will obtain the Poisson equation
in the presence of a minimal length as follows
\begin{equation}
(1-2(\hbar\sqrt{\beta})^{2}\nabla^{2})\nabla^{2}\phi(\textbf{x})=-\frac{\rho(\textbf{x})}{\varepsilon_{0}}.
\end{equation}
The modified Poisson equation (22) has been introduced earlier by
Tkachuk in Ref. 23 with a different approach. The modified Poisson
equation (22) is equivalent to the following two equations
\begin{eqnarray}
(1-a^{2}\nabla^{2})\nabla\cdot\textbf{E}(\textbf{x}) &=&
\frac{\rho(\textbf{x})}{\varepsilon_{0}},\\
\nabla\times\textbf{E}(\textbf{x}) &=& 0,
\end{eqnarray}
where $a := \hbar\sqrt{2\beta}$. Equations (23) and (24) are
fundamental equations of Podolsky's electrostatics [24,25], and $a$
is called Podolsky's characteristic length [26-30].\\
Now, we want to obtain the total potential energy of an
electrostatic field in the presence of a minimal length. Using the
linear higher order equations (22) and (23) together with Eq.
(24), the general expression for the total potential energy
becomes
\begin{eqnarray}
U &=&
\frac{1}{2}\int\rho(\textbf{x})\phi(\textbf{x})d^{3}\textbf{x}\nonumber
\\
&=&
-\frac{1}{2}\varepsilon_{0}\int\phi(\textbf{x})[\nabla^{2}\phi(\textbf{x})-(\hbar\sqrt{2\beta})^{2}\nabla^{2}\nabla^{2}\phi(\textbf{x})]d^{3}\textbf{x}
\nonumber \\
&=& \frac{1}{2}\varepsilon_{0}\int
\{\textbf{E}^{2}-(\hbar\sqrt{2\beta})^{2}[(\nabla\cdot
\textbf{E})^{2}+2\textbf{E}\cdot\nabla^{2}\textbf{E}]\}d^{3}\textbf{x}
\nonumber \\
&=& \frac{1}{2}\varepsilon_{0}\int[\textbf{E}^{2}+
(\hbar\sqrt{2\beta})^{2}(\nabla\cdot\textbf{E})^{2}]d^{3}\textbf{x},
\end{eqnarray}

where we assumed that $\textbf{E}\nabla\cdot\textbf{E}$ falls off
faster than $|\textbf{x}|^{-2}$ as $|\textbf{x}|\rightarrow\infty$.
According to Eq. (25) the energy density of an electrostatic field
in the presence of a minimal length is given by
\begin{equation}
u=\frac{1}{2}\varepsilon_{0}\textbf{E}^{2}+\frac{1}{2}\varepsilon_{0}(\hbar\sqrt{2\beta})^{2}(\nabla\cdot\textbf{E})^{2}.
\end{equation}
The term $
\frac{1}{2}\varepsilon_{0}(\hbar\sqrt{2\beta})^{2}(\nabla\cdot\textbf{E})^{2}$
in Eq. (26) shows the effect of minimal length corrections. Using
the similarity between Podolsky's electrostatics and our modified
electrostatics the electrostatic potential $\phi(\textbf{x})$ for
a point charge $q$ with charge density
$\rho(\textbf{x})=q\delta(\textbf{x})$ can be written as
\begin{equation}
\phi(\textbf{x})=\frac{q}{4\pi\varepsilon_{0}|\textbf{x}|}\left(1-e^{-\frac{|\textbf{x}|}{\hbar\sqrt{2\beta}}}\right).
\end{equation}
In contrast with the usual Maxwell's electrostatics the
electrostatic potential (27) at the origin has a finite value
$\phi(0)=\frac{q}{4\pi\varepsilon_{0}\hbar\sqrt{2\beta}}.$ Now,
using Eq. (27) together with
$\textbf{E}(\textbf{x})=-\nabla\phi(\textbf{x})$ the electric field
due to a point charge $q$ is given by
\begin{equation}
\textbf{E}(\textbf{x})=
\frac{q}{4\pi\varepsilon_{0}|\textbf{x}|^{2}}\left[1-(1+\frac{|\textbf{x}|}{\hbar\sqrt{2\beta}})e^{-\frac{|\textbf{x}|}{\hbar\sqrt{2\beta}}}\right]\frac{\textbf{x}}{|\textbf{x}|}.
\end{equation}
If we substitute (28) into (25) and performing the integration, we
will obtain the total potential energy of a point charge as
$U=\frac{q^{2}}{8\pi\varepsilon_{0}(\hbar\sqrt{2\beta})}.$ By
using Eq. (28), we obtain the modified Coulomb's law for the
electrostatic interaction between a test charge $q_{0}$ and the
point charge $q$ as follows
\begin{eqnarray}
\textbf{F}(\textbf{x})&=& q_{0}\textbf{E}(\textbf{x})\nonumber \\
&=&
\frac{q_{0}q}{4\pi\varepsilon_{0}|\textbf{x}|^{2}}\left[1-(1+\frac{|\textbf{x}|}{\hbar\sqrt{2\beta}})e^{-\frac{|\textbf{x}|}{\hbar\sqrt{2\beta}}}\right]\frac{\textbf{x}}{|\textbf{x}|}.
\end{eqnarray}
In the limit $\hbar\sqrt{2\beta}\rightarrow0$, the modified
Coulomb's law in (29) smoothly becomes the usual Coulomb's law,
i.e.,
\begin{equation}
\lim_{\hbar\sqrt{2\beta}\rightarrow
0}\textbf{F}(\textbf{x})=\frac{q_{0}q}{4\pi\varepsilon_{0}|\textbf{x}|^{3}}\
\textbf{x}.
\end{equation}
\section{Finding the Upper Bound on the Isotropic Minimal Length in Modified Electrostatics}
If we substitute $\beta'=2\beta$ into Eq. (6), we will obtain the
isotropic minimal length in three spatial dimensions as follows
\begin{equation}
(\Delta X_{i})_{min}=\hbar\sqrt{5\beta}\qquad,\qquad \forall\,
i\epsilon\{1,2,3\}.
\end{equation}
According to Eq. (31) the isotropic minimal length can be expressed
in terms of Podolsky's characteristic length $a=\hbar \sqrt{2\beta}$
as
\begin{equation}
(\triangle X_{i})_{min}=\frac{\sqrt{10}}{2}a\qquad,\qquad \forall\,
i\epsilon\{1,2,3\}.
\end{equation}
Now we are ready to estimate the classical and quantum upper bounds
on the isotropic minimal length in modified electrostatics.
\subsection{A Classical Upper Bound on the Isotropic Minimal Length}
In Ref. 28, Accioly and Scatena have obtained a classical bound
on the Podolsky's characteristic length $a$. This classical bound
on $a$ was found using the data from a very accurate experiment
carried out by Plimpton and Lawton [31] to test the Coulomb's law
of force between charges. According to Refs. 28 and 30 the
classical upper limit on $a$ is
\begin{equation}
a\,\leq \, 5.1\times 10^{-10}\,m.
\end{equation}
Using (33) in (32), we obtain the following classical upper bound
for the isotropic minimal length
\begin{equation}
(\triangle X_{i})_{min}\,\leq\, 8.06\times 10^{-10}\, m.
\end{equation}
\subsection{A Quantum Upper Bound on the Isotropic Minimal Length}
In a series of papers, Accioly et al. [26,28,30] obtained a
quantum bound on the Podolsky's characteristic length $a$ by
computing the anomalous magnetic moment of the electron in the
framework of Podolsky's theory. The quantum upper limit on $a$ is
[26,28,30]
\begin{equation}
a\,\leq\,4.7\times10^{-18}\,m.
\end{equation}
Using (35) in (32), we obtain the following quantum upper bound for
the isotropic minimal length
\begin{equation}
(\triangle X_{i})_{min}\,\leq\,7.4\times 10^{-18}\,m.
\end{equation}
Note that the classical upper bound for the isotropic minimal
length in our modified electrostatics is about eight orders of
magnitude larger than the quantum upper bound, i.e.,
\begin{equation}
(\triangle X_{i})_{min-classical}\,\sim \,10^{8}(\triangle
X_{i})_{min-quantum}\,.
\end{equation}
\section{Conclusions}
After 1930, many theoretical physicists have attempted to
introduce a minimal observable length into quantum field theory
[32,33]. The idea of a minimal observable distance received a
great attention in the physics community when it was developed in
details by Heisenberg and March [32,33]. The hope was that the
introduction of such a minimal observable distance would
eliminate divergences that appear in quantum field theories.
Nowadays, we know that the existence of a minimal observable
length leads to a generalized uncertainty principle. An immediate
consequence of the generalized uncertainty principle is a
modification of position and momentum operators according to Eqs.
(12) and (13) for $\beta'=2\beta$. We have shown that the
Lagrangian formulation of an electrostatic field with a charge
density in the presence of a minimal observable length leads to a
fourth-order Poisson equation. Also, we proved that there is a
similarity between electrostatics in the presence of a minimal
length and Podolsky's electrostatics. Another interesting
property of electrostatics in the presence of a minimal length is
that the classical self-energy of a point charge becomes a finite
value.
\\  Now, let us compare the classical and quantum upper bounds on
the isotropic minimal length in this work with the results of
Refs. 34-36. In Ref. 34 the motion of neutrons in a gravitational
quantum well has been studied by Brau and Buisseret and it was
shown that $(\triangle X_{i})_{min}\leq 2.41\times10^{-9}\, m$.
In Ref. 35 Nouicer has studied the Casimir effect in the presence
of a minimal length and obtains $(\triangle X_{i})_{min}\leq
15\times10^{-9}\, m$. The classical upper limit in Eq. (34) is
compatible with the results of Refs. 34 and 35. In Ref. 36 it was
deduced that the upper bound of the minimal length ranges from
$2.4\times 10^{-17} \, m$ to $3.3 \times 10^{-18} \, m$. The
quantum upper limit on the isotropic minimal length in Eq. (36)
is near to the results of Ref. 36. It is necessary to note that
the quantum upper bound on the isotropic minimal length in this
paper, i.e., $7.4\times10^{-18}\, m$ is also near to the
electroweak length scale $(\ell_{electroweak}\sim 10^{-18}\, m)$.
Recently, Smailagic and Spallucci have proposed a novel way to
formulate quantum field theory in the presence of a minimal
length [37]. Using Smailagic-Spallucci approach, Gaete and
Spallucci introduced a $U(1)$ gauge field with a non-local
kinetic term as follows
\begin{equation}
{\cal L}= - \frac{1}{4\mu_{0}}F_{\mu\nu}\;e^{\theta
\partial_{\alpha}\partial^{\alpha}}\;F^{\mu\nu}-J^{\mu}A_{\mu}\,,
\end{equation}
where $\sqrt{\theta}$ is the minimal length of model [38]. The
authors of Ref. 38 have shown in the Appendix A of their paper
that in the electrostatic case and to first order in $\theta$,
the non-local Lagrangian density (38) leads to the following
modified Gauss's law
\begin{equation}
(1-\theta\nabla^{2})\nabla\cdot\textbf{E}(\textbf{x})=\frac{\rho(\textbf{x})}{\varepsilon_{0}}\,.
\end{equation}
A comparison between equations (23) and (39) shows that there is
an equivalence between the Gaete-Spallucci electrostatics to
first order in $\theta$ and the electrostatics with a minimal
observable length. It should be noted that the electrostatics with
a minimal observable length in this study is only correct to the
first order in deformation parameter $\beta$, while the
Gaete-Spallucci electrostatics is valid to all order in $\theta$.

\section*{Acknowledgments}
We would like to thank the referees for their useful comments and
for bringing Ref. 38 to our attention.


\begin{thebibliography}{11}
\bibitem{P1}
B. Bolen and M. Cavaglia, Gen. Rel. Grav. \textbf{37}, 1255
(2005); A. F. Ali, S. Das and E. C. Vagenas, Phys. Lett. B
\textbf{678}, 497 (2009); A. F. Ali, S. Das and E. C. Vagenas,
arXiv: 1001.2642.
\bibitem{P2}
M. S. Berger and M. Maziashvili, Phys. Rev. D \textbf{84}, 044043
(2011).
\bibitem{P3}
M. Kober, Phys. Rev. D \textbf{82}, 085017 (2010).
\bibitem{P4}
M. Kober, Int. J. Mod. Phys. A \textbf{26}, 4251 (2011).
\bibitem{P5}
S. Capozziello, G. Lambiase and G. Scarpetta, Int. J. Theor. Phys.
\textbf{39}, 15 (2000).
\bibitem{P6}
A. Kempf, J. Phys. A: Math. Gen. \textbf{30}, 2093 (1997).
\bibitem{P7}
A. Kempf, G. Mangano and R. B. Mann, Phys. Rev. D \textbf{52},
1108 (1995).
\bibitem{P8}
A. Kempf and G. Mangano, Phys. Rev. D \textbf{55}, 7909 (1997).
\bibitem{P9}
S. K. Moayedi, M. R. Setare and H. Moayeri, Int. J. Theor. Phys.
\textbf{49}, 2080 (2010).
\bibitem{P10}
S. K. Moayedi, M. R. Setare, H. Moayeri and M. Poorakbar, Int. J.
Mod. Phys. A \textbf{26}, 4981 (2011).
\bibitem{P11}
M. R. Setare, Phys. Rev. D \textbf{70}, 087501 (2004).
\bibitem{P12}
D. Bouaziz and M. Bawin, Phys. Rev. A \textbf{76}, 032112 (2007).
\bibitem{P13}
S. Das and E. C. Vagenas, Phys. Rev. Lett. \textbf{101}, 221301
(2008).
\bibitem{P14}
S. Das and E. C. Vagenas, Can. J. Phys. \textbf{87}, 233 (2009).
\bibitem{P15}
F. Brau, J. Phys. A: Math. Gen. \textbf{32}, 7691 (1999).
\bibitem{P16}
S. Benczik, L. N. Chang, D. Minic and T. Takeuchi, Phys. Rev. A
\textbf{72}, 012104 (2005).
\bibitem{P17}
M. M. Stetsko and V. M. Tkachuk, Phys. Rev. A \textbf{74}, 012101
(2006).
\bibitem{P18}
M. M. Stetsko, Phys. Rev. A \textbf{74}, 062105 (2006).
\bibitem{P19}
L. N. Chang, D. Minic, N. Okamura and T. Takeuchi, Phys. Rev. D
\textbf{65}, 125027 (2002).
\bibitem{P20}
G. B. Arfken and H. J. Weber, Mathematical Methods for
Physicists, sixth edition, (Academic Press, 2005).
\bibitem{P21}
J. Magueijo, Phys. Rev. D \textbf{73}, 124020 (2006).
\bibitem{P22}
N. Moeller and B. Zwiebach, J. High Energy Phys. \textbf{10}, 034
(2002).
\bibitem{P23}
V. M. Tkachuk, J. Phys. Stud. \textbf{11}, 41 (2007).
\bibitem{P24}
B. Podolsky, Phys. Rev. \textbf{62}, 68 (1942).
\bibitem{P25}
B. Podolsky and P. Schwed, Rev. Mod. Phys. \textbf{20}, 40 (1948).
\bibitem{P26}
A. Accioly and H. Mukai, Nuovo Cimento B \textbf{112}, 1061
(1997).
\bibitem{P27}
A. Accioly and H. Mukai, Braz. J. Phys. \textbf{28}, 35 (1998).
\bibitem{P28}
A. Accioly and E. Scatena, Mod. Phys. Lett. A \textbf{25}, 269
(2010).
\bibitem{P29}
A. Accioly, Am. J. Phys. \textbf{65}, 882 (1997).
\bibitem{P30}
A. Accioly, P. Gaete, J. Helayel-Neto, E. Scatena and R. Turcati,
arXiv: 1012.1045; A. Accioly, P. Gaete, J. Helayel-Neto, E.
Scatena and R. Turcati, Mod. Phys. Lett. A \textbf{26}, 1985
(2011).
\bibitem{P31}
S. Plimpton and W. Lawton, Phys. Rev. \textbf{50}, 1066 (1936).
\bibitem{P32}
B. Carazza and H. Kragh, Am. J. Phys. \textbf{63}, 595 (1995).
\bibitem{P33}
T. G. Pavlopoulos, Phys. Rev. \textbf{159}, 1106 (1967).
\bibitem{P34}
F. Brau and F. Buisseret, Phys. Rev. D \textbf{74}, 036002 (2006).
\bibitem{P35}
Kh. Nouicer, J. Phys. A: Math. Gen. \textbf{38}, 10027 (2005).
\bibitem{P36}
C. Quesne and V. M. Tkachuk, Phys. Rev. A \textbf{81}, 012106
(2010).
\bibitem{P37}
A. Smailagic and E. Spallucci, J. Phys. A: Math. Gen.
\textbf{36}, L517 (2003); A. Smailagic and E. Spallucci, J. Phys.
A: Math. Gen. \textbf{36}, L467 (2003); A. Smailagic and E.
Spallucci, J. Phys. A: Math. Gen. \textbf{37}, 7169 (2004).
\bibitem{P38}
P. Gaete and E. Spallucci, J. Phys. A: Math. Theor. \textbf{45},
065401 (2012).





\end{thebibliography}
\end{document}